# WHAT ARE THE SUITABLE INSTRUCTIONAL STRATEGY AND MEDIA FOR STUDENT LEARNING STYLES IN MIDDLE SCHOOLS?

Raditya Bayu Rahadian[1] and C. Asri Budiningsih[2]

Department of Instructional Technology, Postgraduate Program, Yogyakarta State University, Yogyakarta, Indonesia

*ABSTRACT*

*This study aims to find out what instructional strategy and media more properly based on learning styles in middle schools. For this purpose, 307 students of $7^{th}$ to $9^{th}$ grade from three middle schools were the respondents. Index Learning Styles was used to know the student learning styles based on Felder-Silverman learning styles. The students were also asked with the questions to know what instructional strategy and media they like much. The results show instructional strategies that could be applied in middle schools are; question and answer methods, student presentations, games and simulations, lectures, problem solving based learning, role playing, and panel discussions. Instructional media that could be applied are; pictures, graphics, videos, simulations, online group learning, newspapers, chatting/messenger, powerpoint slides, computer animations, book/e-book, magazines, audio recorded, and email. This is than checked off with the combinations of student learning style to get the most suitable one.*

## 1. INTRODUCTION

Student characteristic has known as the most important variabel in instructional design [1], [2]. This variable influence instructional strategy and media selecting for learning. Students have individual characteristics that facilitate the degree of success they have with particular intervention programs [3]. Different ways of learning make different strategy and media as learning materials. Why? The different ways of doing the course helps in raising the learning abilities of the student. The teacher displays the information in different ways, with different resources, making the learning process easier due to the fact that some people are more receptive to some kind of information than the others [4].

Student learning styles is one of the most influence student characteristics [1], [5]–[8], but in fact not well applied yet by the teachers in the classroom. By the pre investigation has known that teachers especially in middle school doesn't understand how make a good method for implement the result of student learning style practically. Therefore, instructional implementation going to draging on and unmeaningful for the student. With the result that, need an effort to give a guidance for the teachers in middle school to connecting the selected preferences instructional strategy and media based on the student learning styles.

With regard to the purpose, this paper begins with a literatur review in Section 2. This is then followed by a review of the methodology used for conducting this study in Section 3. Section 4 presents the result of the study. The last part of the paper includes the conclusion and recommendation of the study.

## 2. LITERATURE REVIEW

### 2.1. LEARNING STYLES

There is no one single method of learning. There are many and works best depends on the task, the context and student personality. Student will be a more effective learner if aware of the range of possible learning methods, and know when to apply it for the best results [9]. Oftentimes, the decision of instructional design concern with too much learning matters, so that omitting consideration on how student can learn it's matters, and selecting the method to achieve instructional goals. Even if the student himself oftenly unrealized with their conditions, preferences, and strenghtness for better learning. While the key of effective and efficiencies of instructional depend on how student care with their preferences of learning that highly connected with learning styles.

Learning style is a kind of student characteristics that include on the instructional condition variables. The framework connections between all instructional variables describe on Figure 1.

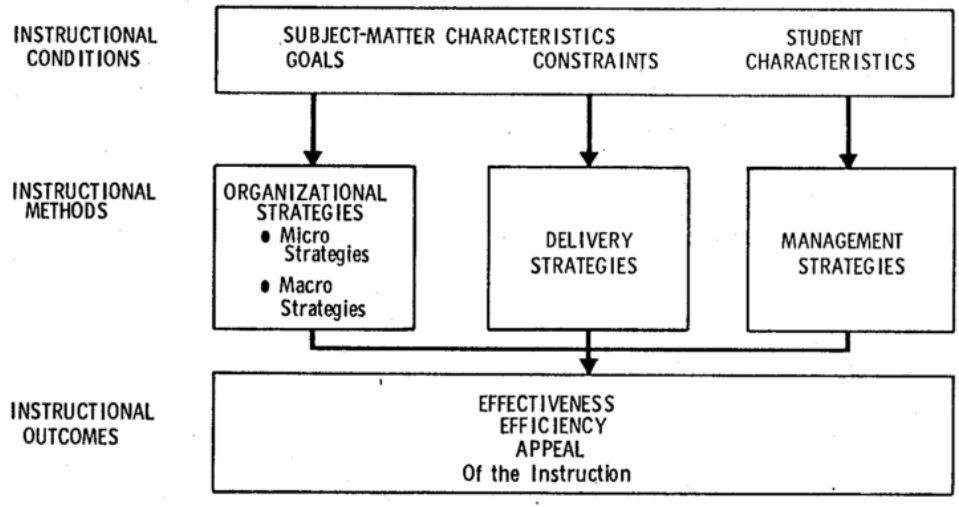

Figure 1. Frameworks connection between all instructional variables [1]

As describe on Figure 1, instructional condition variables influence other variables in straight down arrow. This is mean that the condition variables must consider first before designing an instructional program to achieve desire outcome. This variables must accept without any question and could not manipulated in instructional. In addition to design an instructional, condition variables influence selecting methods of learning. Condition variables are always be independent, even in descriptive theory or in prescriptive theory of instruction [1]. Therefore analyzing this variables have to correctly and objectively so that convenient with the desirable instructional goals.

Theoretically the student have differences on personality, general potencies, and knowledge in one or two subject matter [10], as clearly describe on Figure 2.

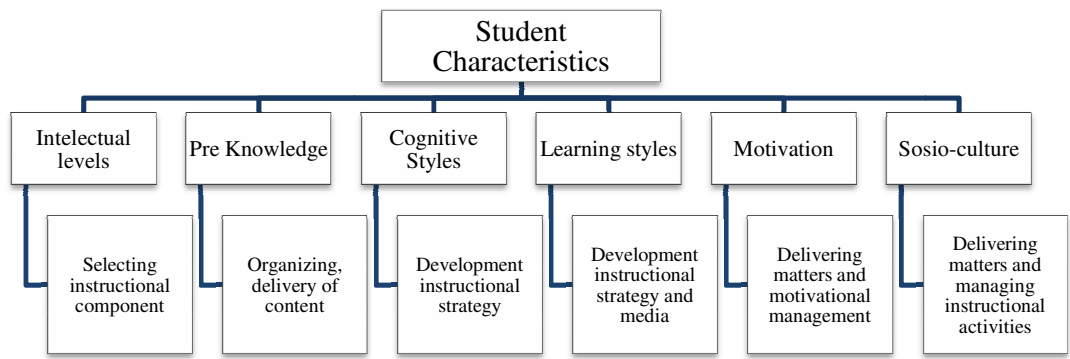

Figure 2. Student characteristics influence the selection of instructional strategies

The variety of student characteristics as describe on Figure 2 influence the selection of other instructional method variables. For example the information of intelectual levels influence the selection of instructional component. The cognitive development connected with the student ability for receive and determining truth and preferences to gather evidence judgments, and openness to change if new evidence is forthcoming [5]. The information of cognitive styles helping teachers to supporting media or learning experiences that match with the level of students cognitive developmental, so that student feel comfort and equal with instructional activities on progress. So with the other kind of student characteristics have different effect in instructional strategies such as selection and organizationing of matters, strategies, media, and other learning resources.

Learning styles has been choosen as one of the student characteristics in this study due to it's big effect for developmenting instructional strategy and media. Knowing student learning styles useful not only for the teachers in addition to design instructional with excellent quality, but also for the students to more understanding their strength for learning optimally. There is the benefit of knowing learning styles for students:

a.  Help student understand their own likely approach to learning opportunities, and perhaps how to use that basic approach better [11].
b.  Increase student learning ability [11].
c.  Povide some guidance for helping student develop self-direction and self-assessment [12].
d.  Maximize the benefit of learning styles their own to support range of styles to learn individually [11], [13].

While the benefit of knowing learning styles for teachers are:

a.  Help teacher deliver the different choices apropriate with student learning styles [11]–[13].
b.  Identify the overall assessment strategy – for example, self-report, observation [12].
c.  Develop individual programmes for the children in the class in accordance with the curriculum content and objectives [12].
d.  Plan how the content can be differentiated in terms of presentation [12].
e.  Identify resources that will be necessary to support the range of styles [12].
f.  Identify and plan the classroom environment that can incorporate the range of styles [12].

As describe first, learning style influence development of the instructional strategy and media. Felder & Silverman stated that student learning styles according to where they fit on a number of scales pertaining to the ways they receive and process information [14]. Other expert give the definition of learning styles as the way in which each person begins to concentrate on, process,

27

and retain new and difficult information. Concentration occurs differently for different people at different times [13]. Hartley defines learning styles as the strategies students adopt when studying. Different strategies can be selected by learners to deal with different tasks [15]. In different manner, learning styles can be interpreted on two ways, (a) differential preferences for processing certain types of information or (b) for processing information in certain ways [16].

More comprehensive definition of learning styles has been defined by Alan Pritchard; (a) a particular way in which an individual learns; (b) a mode of learning – an individual's preferred or best manner(s) in which to think, process information and demonstrate learning; (c) an individual's preferred means of acquiring knowledge and skills; and (d) habits, strategies, or regular mental behaviours concerning learning, particularly deliberate educational learning, that an individual displays [17].

Some definitions has been explained that basically someone have strenght, character, and specific preferences to receive and process information while at the same time more like certain information when learning than others. This is show that everyone have a special way for learning that makes different look to other ones. Undoubtable that learning styles refers to the fact that every person has its own method or set of strategies when learning [4].

There are several models of learning style. Some of them are Kolb learning styles [5], [15], Honey & Mumford models [15], [17], V-A-R-K models [17], Myers-Briggs Type Indicator (MBTI) models [5], [17], [18], Felder-Silverman models [5], [6]. On this study selected Felder-Silverman models for the following reasons:

a. It is one strong proponent of learning styles and their influence on both learning and the design of instruction [6].
b. It has been successfully implemented in recent study about the adaptation of student learning styles to the use of instructional matters and strategies [4], [19]–[24].
c. It is user friendly even if for student him/her self and the result is easy to interpretate [4], [25].
d. The Index Learning Styles (ILS) is trough a deep and long validation so that proper to defining student learning styles [26]–[31].
e. The Index Learning Styles (ILS) is available free to individuals and instructors who wish to use it for teaching and research on their own classes [32].

**2.2. FELDER-SILVERMAN LEARNING STYLES MODEL**

The power of Felder-Silverman learning styles model is adressed on the combination of four dimension learning styles, perception, entry channel, processing, and understanding. Each dimensions have two type that contrary another. Felder-Silverman model of learning styles is based on the answer of four questions [4], [5], [14]:

a. What kind of information does the student tend to receive (perception):
   1) Sensitive (**S**): from external agents (places, sounds, physical sensations)
   2) Intuitive (**I**): from internal agents (memories, possibilities, ideas, insights)
b. Through which sensorial channel do the students tend to receive information more effectively (entry chanel)?
   1) Visual (**Vi**): images, picture, flowchart, diagrams, graphics, demostrations
   2) Verbal (**Ve**): spoken words, sounds, written text
c. How is the information processed (processing)?
   1) Actively (**A**): physical activities and discussions
   2) Reflectively (**R**): introspection

d.  How does the student make progress (understanding)?
    1) Sequentially (**S**): continuous step in a logical progression of incremental steps
    2) Globally (**G**): leaps in large "big pictures" jumps and an integral approaches

Table 1 shows the conditions of content and learners must have paid attention according to the type of learning styles.

Table 1. The conditions of content and learners according to the type of learning styles

| Dimensions of learning styles | Type of learning styles | The content | The learners |
|---|---|---|---|
| Processing | Active (A) | must be applicable | do not learn much in situations that require them to be passive, work well in groups, tend to be experimentalists |
| | Reflective (R) | must be related with experiences | do not learn much in situations that provide no opportunity to think about the information being presented, work better by themselves or with at most one other person, tend to be theoreticians |
| Perception | Sensitive (S) | concrete, practical, immediate connection with the real word | oriented toward facts and hands-on procedures |
| | Intuitif (I) | innovative, oriented to theory and meanings, avoiding repetitive methods | comfortable with abstractions, rapid and innovative problem solvers |
| Entry channel | Visual (Vi) | heavy on visual components, information gathering must use visual representations | actions to visualize, remember best what they see, will probably forget something is simply said to them |
| | Verbal (Ve) | lot of oral and textual components, the information gathering must use textual representations | remember much of what they hear and more of what they hear and then say |
| Understanding | Sequential (Se) | must be written orderly, step by step, associated to the problems being solved | tend to think in a linear manner and are able to function with only partial understanding of material they have been taught |
| | Global (G) | must be written in big leaps, suddenly and almost randomly | can solve complex problems quickly and put things together in an innovative way but may have difficulties to explain how they did it |

Index Learning Styles (ILS) is used to categorize the type of learning style. It is a fourty four item forced choice instrument to assess preferences on the four scales of the Felder-Silverman model. Each question have two choices that determine the strenght or preferences of their learning styles [33]. Each of learning style dimensions have specific implication on selecting instructional strategy and media. Perception influence selecting of matter. Entry channel influence on presentation. Processing influence student participation. Understanding influence view of student perspective learning [14].

The combination of learning style is composed of sixteen type of learning style as clearly describe on Table 2.

Table 2. Combinations of Felder-Silverman learning styles

| No | Combinations of Felder-Silverman Learning Styles | Initials |
|---|---|---|
| 1 | Active, Sensitive, Visual, Sequential | A-S-Vi-Se |
| 2 | Active, Sensitive, Visual, Global | A-S-Vi-G |
| 3 | Active, Intuitive, Visual, Sequential | A-I-Vi-Se |
| 4 | Active, Intuitive, Visual, Global | A-I-Vi-G |
| 5 | Active, Sensitive, Verbal, Sequential | A-S-Ve-Se |
| 6 | Active, Sensitive, Verbal, Global | A-S-Ve-G |
| 7 | Active, Intuitive, Verbal, Sequential | A-I-Ve-Se |
| 8 | Active, Intuitive, Verbal, Global | A-I-Ve-G |
| 9 | Reflective, Sensitive, Visual, Sequential | R-S-Vi-Se |
| 10 | Reflective, Sensitive, Visual, Global | R-S-Vi-G |
| 11 | Reflective, Intuitive, Visual, Sequential | R-I-Vi-Se |
| 12 | Reflective, Intuitive, Visual, Global | R-I-Vi-G |
| 13 | Reflective, Sensitive, Verbal, Sequential | R-S-Ve-Se |
| 14 | Reflective, Sensitive, Verbal, Global | R-S-Ve-G |
| 15 | Reflective, Intuitive, Verbal, Sequential | R-I-Ve-Se |
| 16 | Reflective, Intuitive, Verbal, Global | R-I-Ve-G |

Every combination is assosiated with instructional strategy and media that much suitable by student him self. The instructional strategy and media that much prefer to each type of learning style can be made as a references for the teachers when prepare instructional activity, especially for middle schools.

### 2.3. INSTRUCTIONAL STRATEGIES

Strategy on instructional is about the special way that teacher gives to the student to help facilitating their circumstantial understanding while learning an information. This is connected with planning, programming, elaborating, and determining the attainment of learning matters.
In middle school, student need autonomy and the ability to maker their own decisions. So that teacher must provide more group work, with an emphasis on effort and improvement. Teachers can also provide students more choices to allow for greater autonomy [34]. Applications of instructional strategies in middle school then must be done in addition to increasing the student abilities on perceiving, analizing, expressing opinion, drawing hypothesis, finding solutions, and exploring knowledge autonomously.

By the recent study there are kind of instructional strategies that could be applied: games and simulations, problem solving based learning, role playing, lectures, student presentations, panel discussions, brainstorming, case study, question and answer, and project design [4], [35], [36].

## 2.4. INSTRUCTIONAL MEDIA

A media is a means of communication and source of information. It is refers to anything that carries information between a source and a receiver. When they provide messages with an instructional purpose, then these are considered as an instructional media [10]. The main purpose of instructional media is to facilitate communication and learning.

Six basic concept of media are text, audio, visuals, motion media, manipulatives (objects), and people [10]. The widely kinds of media are still picture, pictures/cartoon, chart/graphics, posters, audio recorded, audio conference, online group learning, chatting/messenger, e-mail, computer animations, simulations, magazines, newspapers, book/e-book, hypertext/web page, slideshows, internet research, tutorials, live video recorded, and video conference [4], [10].

According to the widely kinds of instructional strategy and media need a limitation as suitable for the purpose of the study. The limitation is based on the strategy and media use in schools where this study take place. So that only real instructional strategy and media are applied really in the class is used in this study that might perhaps different from what described first.

## 2.5. RELATED WORKS

Recent investigations about adaptation models of Felder-Silverman's learning styles try to integrate instructional strategies and also media based on Felder-Silverman learning style. The model has described that learning styles could made differences on giving vary of instructional strategies and media in addition to give more effectivenes and efficiences learning.

Table 3. Adaptation models of Felder-Silverman learning styles

| Author(s) | On based adaptation models | Applications |
| --- | --- | --- |
| Franzoni & Assar, 2009 [4] | types of electronic media are more suitable to a particular learning styles | e-learning |
| Dorca et.al, 2016 [19] | recommendation learning objects considering students learning styles | e-learning |
| Lopes & Fernandes, 2009 [35] | some instructional actions are more appropriate for a learning style than others | e-learning |
| Kowalski & Kowalski, 2013 [21] | Student learning styles effect in interactive simulations are coupled with real-time formative assessment | Both e-learning and real classroom |
| Li, 2015 [22] | identify specific learning style preferences that might be favourable to accepting the proposed pedagogy | Integrate a collaborative technology in classroom |
| Popescu, Badica, & Moraret, 2010 [23] | Provide an extensive review of existing learning style-based adaptive educational systems | e-learning |
| Psycharis, Botsari, & Chatzarakis, 2014 [36] | examine the impact of the computational experiment approach, learning styles, epistemic beliefs, and engagement with the inquiry process on the learning performance | Computer based learning |
| Rajper at.al, 2016 [24] | Automatic detection of E-learners' learning styles is an important requirement for personalized e-learning | e-learning |

This review shows that Felder-Silverman learning style have been applied on instructional with several adaptations to find the finest ways of learning. This is including the application of

31

International Journal on Integrating Technology in Education (IJITE) Vol.6, No.4, December 2017

instructional system, media, and learning approaches. However, the adaptations seems to be limited, while it is depending on the specific goals to each study. It is like the appropriate level of student and kind of learning applications. Except Franzoni & Assar [4] adaptations that appropriate for at least high school students, they are more appropriate for college levels. Also most of them is used for e-learning and computer based learning, not for the real classroom, except Kowalski & Kowalski [21] and Li [22]. So that need more study to find the finest way of adaptations the Felder-Silverman learning styles to apply in middle schools, especially to help teachers selecting the instructional strategy and media based on the real student needs. In this sense, this work is new and significantly different from the previous study done in the field.

## 3. METHODOLOGY

### 3.1. CONTEXT AND PARTICIPANTS

Student learning styles is one of the most influence student characteristics [1], [5]–[8], but in fact not well applicated yet by teachers in the classroom. By the pre investigation has known that teachers in middle schools doesn't understand what the suitable methods will be given to the students that match with their learning styles. This study aims to find out the middle schools students most suitable instructional strategy and media according to their preferences of learning styles. This investigations based on their own likely strategy and media that mostly applicated in the classrooms. The population used in this study included 307 middle schools students of $7^{th}$ to $9^{th}$ grade from three middle schools.

### 3.2. PROCEDURE

This study took place after end of semester exam before the break so that student can remember all about instructional they have been done before for all semester without worrying the exam. Index of Learning Styles [33] was used as a questionnaire to find out the students learning styles. List of questions was used to know what are the instructional strategy and media that have been used in classrooms by the teachers. The students were also asked with the questions to know their likely much with the instructional strategy and media have been used before. The answer will be classified by the proper of student learning style, strategy and media their prefer in instructional programs.

### 3.3. RESEARCH DESIGN

Students fill in the ILS questionnaire to know their type of learning styles. Students answer the questions about instructional strategy and media that have been used in classrooms by the teachers. This information is used to know what instructional strategy and media use in middle schools. The students were also asked with the questions to know the instructional strategy and media have been used before in classrooms as they like much. The design of study as shown in Figure 3.

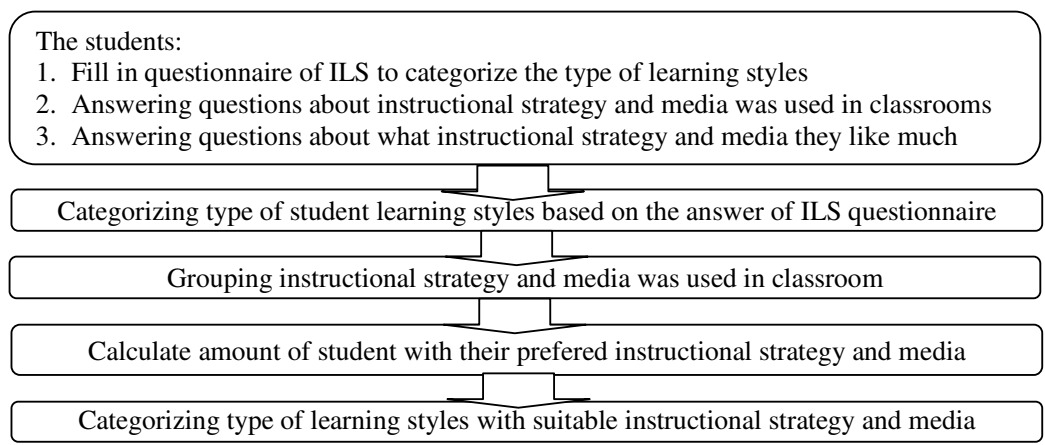

Figure 3. Design of study

## 4. RESULTS

### 4.1. RESULT OF THE QUESTIONNAIRE

The questionnaire used Index of Learning Styles (ILS). It is an instrument to assess preferences on the four scales of the Felder-Silverman model. All 307 participants responded to the questionnaire. The result as shown in Table 4.

Table 4. Result of ILS questionnaire

| Dimensions of learning styles | Type of learning styles | Amount of student | Percentage |
|---|---|---|---|
| Processing | Active | 220 | 72% |
| | Reflective | 87 | 28% |
| Perception | Sensitive | 242 | 79% |
| | Intuitif | 65 | 21% |
| Entry channel | Visual | 157 | 51% |
| | Verbal | 150 | 49% |
| Understanding | Sequential | 194 | 63% |
| | Global | 113 | 37% |

The results show that in the processing dimension, students are more active than reflective. In the perception dimension there are most significant difference, where 79% students came out to be sensitive and only 21% were intuitif. In the entry channel dimension, it was found that the students are almost equal both visual and verbal. Finally, in the understanding dimension, it shows that they are more sequential than global. As a result generally, the predominant combination for each dimension style of the study is Active-Sensitive-Visual-Sequential (A-S-Vi-Se). After categorizing the type of learning styles, next step is categorizing each type to the combinations of learning styles. Is the A-S-Vi-Se still predominating the combinations? The result will be known in Table 5.

Table 5. Result of Felder-Silverman learning styles combinations

| No. | Learning Styles Combinations | Amount of student | Percentage |
|---|---|---|---|
| 1 | A-S-Vi-Se | 57 | 19% |
| 2 | A-S-Vi-G | 30 | 10% |
| 3 | A-I-Vi-Se | 12 | 4% |
| 4 | A-I-Vi-G | 13 | 4% |
| 5 | A-S-Ve-Se | 49 | 16% |
| 6 | A-S-Ve-G | 40 | 13% |
| 7 | A-I-Ve-Se | 13 | 4% |
| 8 | A-I-Ve-G | 6 | 2% |
| 9 | R-S-Vi-Se | 20 | 7% |
| 10 | R-S-Vi-G | 13 | 4% |
| 11 | R-I-Vi-Se | 12 | 4% |
| 12 | R-I-Vi-G | 0 | 0% |
| 13 | R-S-Ve-Se | 18 | 6% |
| 14 | R-S-Ve-G | 9 | 3% |
| 15 | R-I-Ve-Se | 13 | 4% |
| 16 | R-I-Ve-G | 2 | 1% |

As could be seen in Table 5, there are all learning style combinations except combinations number 12, R-I-Vi-G. The A-S-Vi-Se is still predominating the combination of learning styles in this study.

### 4.2. RESULT OF THE QUESTIONS ANSWERED

Student was asked with four items about the instructional strategies and media in their own classroom and what they prefer much about it. Number of participants is 307 students. It is same students who fill in the ILS questionnaire above.

**Item number 1:** What was the instructional strategies used in your classroom by teachers? Write three kinds of instructional strategies that most often used by your teachers.
**Item number 2:** According to the answer of item number 1, choose two instructional strategies that you most like to used.

The result of the item number 1 and 2 as shown in Table 6.

Table 6. Instructional strategies was used in classrooms

| No. | Instructional strategies | Used in classrooms | Liked by students |
|---|---|---|---|
| 1 | Question and answer methods | 75% | 59% |
| 2 | Student presentation | 57% | 35% |
| 3 | Games and simulations | 45% | 26% |
| 4 | Lectures | 43% | 27% |
| 5 | Problem solving based | 40% | 24% |
| 6 | Role playing | 21% | 14% |
| 7 | Panel discussions | 19% | 16% |

Item number 3: What was the instructional media used in your classroom by teachers? Write three kinds of instructional media that most often used by your teachers.

Item number 4: According to the answer of item number 3, choose two instructional media that you most like to used?
The result of the item number 3 and 4 as shown in Table 7.

International Journal on Integrating Technology in Education (IJITE) Vol.6, No.4, December 2017

Table 7. Instructional media was used in classrooms

| No. | Instructional media | Used in classrooms | Liked by students |
|---|---|---|---|
| 1 | Pictures | 61% | 53% |
| 2 | Graphics | 32% | 16% |
| 3 | Video | 30% | 23% |
| 4 | Simulations | 23% | 10% |
| 5 | Online group learning | 22% | 14% |
| 6 | Newspapers | 22% | 15% |
| 7 | Chatting/mesengger | 21% | 12% |
| 8 | Powerpoint slides | 19% | 9% |
| 9 | Computer Animations | 16% | 11% |
| 10 | Book/e-book | 16% | 9% |
| 11 | Magazines | 15% | 9% |
| 12 | Audio recording | 13% | 10% |
| 13 | E-mail | 9% | 8% |

Instructional strategies and media were liked by students then checked off with the combinations of student learning styles that had known by the result of ILS questionnaire. This is the final result as the aim of this study as shown in Table 8 and 9.

Table 8. Suitable instructional strategies for the combinations of student learning styles

| No | Learning Styles Combinations | Question and answer | Student presentations | Games and simulations | Lectures | Promblem solving based learning | Role playing | Panel discussions |
|---|---|---|---|---|---|---|---|---|
| 1 | A-S-Vi-Se | 25% | 14% | 26% | 14% | 16% | 5% | 0% |
| 2 | A-S-Vi-G | 19% | 11% | 37% | 11% | 22% | 0% | 0% |
| 3 | A-I-Vi-Se | 25% | 25% | 25% | 17% | 8% | 0% | 0% |
| 4 | A-I-Vi-G | 8% | 8% | 31% | 15% | 23% | 0% | 15% |
| 5 | A-S-Ve-Se | 35% | 14% | 12% | 14% | 11% | 10% | 5% |
| 6 | A-S-Ve-G | 33% | 18% | 25% | 8% | 13% | 3% | 3% |
| 7 | A-I-Ve-Se | 15% | 23% | 8% | 23% | 31% | 0% | 0% |
| 8 | A-I-Ve-G | 20% | 0% | 40% | 0% | 20% | 20% | 0% |
| 9 | R-S-Vi-Se | 15% | 0% | 30% | 35% | 15% | 0% | 5% |
| 10 | R-S-Vi-G | 8% | 31% | 23% | 8% | 15% | 8% | 8% |
| 11 | R-I-Vi-Se | 42% | 0% | 25% | 0% | 17% | 17% | 0% |
| 12 | R-I-Vi-G | 0% | 0% | 0% | 0% | 0% | 0% | 0% |
| 13 | R-S-Ve-Se | 18% | 6% | 29% | 6% | 41% | 0% | 0% |
| 14 | R-S-Ve-G | 50% | 13% | 0% | 13% | 13% | 13% | 0% |
| 15 | R-I-Ve-Se | 38% | 8% | 15% | 15% | 8% | 15% | 0% |
| 16 | R-I-Ve-G | 50% | 50% | 0% | 0% | 0% | 0% | 0% |

Table 9. Suitable instructional media for the combinations of student learning styles

| No | Learning Styles Combination | Pictures | Graphics | Video | Simulations | Online group learning | Newspapers | Chatting/ messenger | Powerpoint slides | Computer animation | Book/e-book | Magazines | Audio recorded | E-mail |
|---|---|---|---|---|---|---|---|---|---|---|---|---|---|---|
| 1 | A-S-Vi-Se | 28% | 7% | 11% | 5% | 6% | 7% | 2% | 5% | 5% | 3% | 3% | 8% | 5% |
| 2 | A-S-Vi-G | 36% | 9% | 20% | 0% | 2% | 7% | 7% | 0% | 5% | 2% | 0% | 2% | 5% |
| 3 | A-I-Vi-Se | 22% | 4% | 17% | 4% | 9% | 4% | 4% | 9% | 4% | 4% | 13% | 0% | 4% |
| 4 | A-I-Vi-G | 29% | 0% | 24% | 5% | 0% | 5% | 10% | 5% | 14% | 5% | 5% | 0% | 0% |
| 5 | A-S-Ve-Se | 23% | 8% | 17% | 8% | 8% | 6% | 8% | 0% | 6% | 2% | 8% | 2% | 3% |
| 6 | A-S-Ve-G | 33% | 2% | 11% | 0% | 11% | 5% | 9% | 7% | 5% | 2% | 2% | 7% | 0% |
| 7 | A-I-Ve-Se | 28% | 12% | 4% | 4% | 4% | 12% | 4% | 4% | 4% | 8% | 0% | 4% | 12% |
| 8 | A-I-Ve-G | 43% | 14% | 0% | 0% | 0% | 14% | 0% | 0% | 0% | 0% | 14% | 0% | 14% |
| 9 | R-S-Vi-Se | 20% | 11% | 11% | 4% | 4% | 7% | 7% | 11% | 7% | 2% | 2% | 9% | 2% |
| 10 | R-S-Vi-G | 24% | 8% | 8% | 4% | 8% | 12% | 8% | 0% | 4% | 4% | 4% | 4% | 0% |
| 11 | R-I-Vi-Se | 36% | 9% | 0% | 5% | 9% | 5% | 9% | 5% | 0% | 5% | 5% | 9% | 5% |
| 12 | R-I-Vi-G | 0% | 0% | 0% | 0% | 0% | 0% | 0% | 0% | 0% | 0% | 0% | 0% | 0% |
| 13 | R-S-Ve-Se | 35% | 3% | 10% | 0% | 10% | 13% | 0% | 0% | 3% | 10% | 3% | 6% | 3% |
| 14 | R-S-Ve-G | 30% | 20% | 0% | 10% | 20% | 10% | 0% | 0% | 0% | 0% | 0% | 0% | 0% |
| 15 | R-I-Ve-Se | 30% | 15% | 5% | 0% | 5% | 15% | 0% | 0% | 10% | 5% | 5% | 0% | 5% |
| 16 | R-I-Ve-G | 67% | 0% | 0% | 0% | 0% | 0% | 0% | 0% | 0% | 0% | 0% | 0% | 33% |

## 5. DISCUSSION AND CONCLUSION

This study aimed to find out what instructional strategy and media more suitable based on student learning styles in middle schools. The goal of this study was also to help teachers selecting the instructional strategy and media based on the real student needs. This is the way to achieve a meaningful learning for students [9]. The different approaches to the kinds of student learning styles would be riched the resources of instructional [11]–[13].

According to the result of the study, the instructional strategies that could be applied in middle schools are question and answer methods, student presentations, games and simulations, lectures, problem solving based learning, role playing, and panel discussions. Although question and answer methods is the predominant strategy than others, lectures still interesting in middle schools. This is show that even student need autonomy and the ability to maker their own decisions [34], they still need the teachers to explained some of the learning content.

The instructional media that could be applied according to the result of the study are pictures, graphics, videos, simulations, online group learning, newspapers, chatting/messenger, powerpoint slides, computer animations, book/e-book, magazines, audio recorded, and email. All the instructional media as written in this study perhaps is the most applied media in the class and be the most liked by student as their prefered learning styles. Hence the range of instructional media become a much according to the most student prefer with.

According to the result, the most suitable instructional strategies and media based on the student learning styles could be used by teachers in middle schools. Big percentage of strategy or media could be considered as the most suitable ones. For example, the combination of learning style A-S-Vi-Se (active, sensitive, visual, sequential) more suitable with games and simulations, questions and answer methods, problem solving based learning, and student presentations, but maybe still need lecture and role playing for some learning content. A-S-Vi-Se is also more suitable with pictures, videos, or graphics. But sometimes maybe still need an audio recorded in case that the weight of visual and verbal learners on this study almost equal.

There is none of the participant who have the R-I-Vi-G (Reflective, Intuitive, Visual, Global) learning style combination. But according to the resemble combinations (R-I-Ve-G, and R-I-Vi-Se) could be considered to use the questions and answer methods, student presentations, and games and simulations as the suitable instructional strategies. And also the use of pictures, email, and chatting/messenger as the suitable instructional media for R-I-Vi-G learning styles combination could be considered on. However need more study to declare the suitable instructional strategies and media for the R-I-Vi-G combination appropriately based on the real data.

The future study should collect data on the relationship of this application with the result of the student learning. This will help better undestanding the suitable instructional strategies and media when apllied in the suitable learning style will give better result for the student or not, and what should be.

## ACKNOWLEDGEMENTS

We would like to express our sincerest thanks and appreciation to Indonesia Endowment Fund for Education (LPDP) as a funder of this research.

## AUTHORS

**Raditya Bayu Rahadian** hold his master's degree in instructional technology from Yogyakarta State University, Yogyakarta, Indonesia. He is an instructional technologist developer in South Bangka Regency, Indonesia since 2010. He is interested in research in the field of educational technology, including instructional technology, educational services, and educational management.

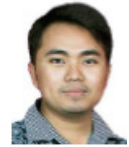

**C. Asri Budiningsih** is a professor in Yogyakarta State University. She holds her Doctor in instructional technology from Malang State University, Indonesia. She is a senior researcher in department of education in Yogyakarta State University. She is interested in research in the field of general educational system, including educational philosophy, educational curricullum, and educational/instructional technology.

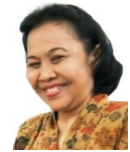